\documentclass[aps,pre,showkays,twocolumn,superscriptaddress,floatfix]{revtex4}

\usepackage{graphicx}
\usepackage{color}

\usepackage{epsfig}
\usepackage{epstopdf}
\usepackage{amsfonts}
\usepackage{amssymb,amsmath}

\begin{document}

\title[Deformed Cornu spiral] {Riccati parametric deformations of the Cornu spiral}

\author{Haret C. Rosu}
\email{hcr@ipicyt.edu.mx}
\affiliation{IPICYT, Instituto Potosino de Investigacion Cientifica y Tecnologica,\\
Camino a la presa San Jos\'e 2055, Col. Lomas 4a Secci\'on, 78216 San Luis Potos\'{\i}, S.L.P., Mexico}

\author{Stefan C. Mancas}
\email{mancass@erau.edu}
\affiliation{Department of Mathematics, Embry-Riddle Aeronautical University, Daytona Beach, FL 32114-3900, USA}

\author{E. Flores-Gardu\~no}
\email{elizabeth.flores@ipicyt.edu.mx}
\affiliation{IPICYT, Instituto Potosino de Investigacion Cientifica y Tecnologica,\\
Camino a la presa San Jos\'e 2055, Col. Lomas 4a Secci\'on, 78216 San Luis Potos\'{\i}, S.L.P., Mexico}


\begin{abstract}
A parametric deformation of the Cornu spiral is introduced.
The parameter is an integration constant which appears in the general solution of the Riccati equation related to the Fresnel integrals.
Argand plots of the deformed spirals are presented and a supersymmetric (Darboux) structure of the deformation is revealed through the
factorization approach.\\


\begin{center} accepted at Zeitschrift f\"ur Naturforschung A (2018) \end{center}

\end{abstract} 

\vspace{2pc}

\maketitle

\section{Introduction} 

One of the most famous spirals with important scientific and technological consequences is Euler's spiral, also known as Cornu's spiral in optics, and also as the clothoid, which means looking-like Clotho, as proposed by note geometer Ces\`aro in 1890.
Schwartzman, in his book ``The Words of Mathematics'' \cite{Steven}, mentions that Clotho was the youngest of the three fates {\it ``moirai"} in ancient Greek mythology. The little sister Clotho was responsible for spinning the thread of human life. Presumably, Ces\`aro was inspired by the resemblance of the spiral to a spinning wheel. However, here, we will call this spiral as Cornu's spiral since it was Cornu who first drew the entire spiral with its two foci, while Euler drew only the positive arm.

Perhaps the simplest mathematical definition of the Cornu spiral $\cal{F}$, is as the Argand plane representation, ${\cal F}=X+iY$, with $X$ and $Y$, the two Fresnel integrals 
\begin{eqnarray}\label{Fes}
	C(z) &=& \int_{0}^{z} \cos \left( \frac{\pi}{2} s^{2} \right) ds~\equiv X, \label{Fes1}\\
	S(z) &=& \int_{0}^{z} \sin \left( \frac{\pi}{2} s^{2} \right) ds~\equiv Y,\label{Fes2}
\end{eqnarray}
which are parametrized by the arclength of the spiral, $s$. In optics, the square modulus $|{\cal F}|^2$ is related to the intensity of light at a given point in diffraction patterns.

On the other hand, geometrically, the Cornu spiral is defined as the curve whose curvature increases linearly with arclength, which means the radius of curvature $\rho (t)$ times the arc length $s(t)$ is constant at each point of the curve. This is represented by the Ces\`aro equation $ \rho (t) = \frac{c^{2}}{s(t)}$, with $c$ any constant. 
Other important property is related to the Fresnel integrals for which $c=\frac{1}{\sqrt{\pi}}$; both approach slowly the point $( \frac{1}{2}, \frac{1}{2} )$ as $s \rightarrow \infty$ in the first quadrant, and because both functions are odd, the curve spirals towards $( - \frac{1}{2}, - \frac{1}{2} )$ in the third quadrant \cite{Fowles}.

Its most immediate technological use is in the layout of civil engineering works (roads, railways, pipelines, among others) as road transitions to join straight sections with curved sections or to connect two circular sections \cite{baas,walton}. This is one of its most important engineering application, since the radius of curvature decreases inversely proportional to the distance traveled on it, and this feature allows the driver a smooth change of trajectory. 
Other applications in which clothoids have been considered are for controlled trajectories of robots \cite{fleury}, and in designing roller coasters \cite{pendrill}, and aesthetic shapes of industrial products \cite{gobit}.

Various generalizations of the Cornu spiral from the viewpoint of different applications can be found in the literature \cite{g1,g2,g3,g4}.
In this communication, we introduce a parametric generalization which can be also considered as a deformation of the Cornu spiral. This is achieved by means of a complex parameter which appears in the general solution of the Riccati equation that corresponds to the Fresnel integrals. In Section 2, we show the reduction of the third order ordinary differential equation (ode) satisfied
by the Fresnel integrals as particular solutions to the corresponding Riccati equation, whose general solution is obtained explicitly. We then write the solution of the third order ode based on the general Riccati solution and present Argand plots of this solution. In Section 3, the similarity with supersymmetric quantum mechanics is emphasized by means of the factorization approach \cite{M84,RMC1,MRO,R98,C01} which is applied to the second-order linear ode that comes into play in the reduction process of Section 2.

\section{From the third order ode to the Riccati equation and back}
We start with the known linear third order ode satisfied by the Fresnel integrals \cite{Wolfram}
\begin{equation}\label{eq1}
z{\rm w}'''-{\rm w}''+\pi^2 z^3 {\rm w}\rq{}=0~,
\end{equation}
which can be reduced by using ${\rm w}\rq{}(z)=v (z)$, with $^\prime=\frac{d}{dz}$,  to obtain
\begin{equation}\label{eq2}
v''-\frac 1 z v'+\pi^2 z^2 v =0.
\end{equation}
Letting $z^2=\zeta$ we obtain the simple harmonic oscillator
\begin{equation}\label{eq3}
\frac{d^2v}{d\zeta^2}+\left(\frac{\pi}{2}\right)^2v=0~.
\end{equation}
Thus, the solution for (\ref{eq2}) is
\begin{equation}\label{eq5}
v(z)=c_1 \cos \left( \frac \pi 2 z^2\right)+c_2 \sin \left( \frac \pi 2 z^2\right)~,
\end{equation}
and by one  integration the solution to (\ref{eq1}) is
\begin{equation}\label{eq7}
{\rm w}(z)=c_1 C(z)+c_2 S(z)+{\rm w}(0)~,
\end{equation}
where $C(z)$, and $S(z)$ are the Fresnel integrals given by (\ref{Fes1}) and (\ref{Fes2}).

On the other hand, using the  logarithmic derivative $y(z)=\frac {v'(z)}{v(z)}$,
(\ref{eq2})  becomes the Riccati equation
\begin{equation}\label{eq10}
y\rq{}+y^2=\frac 1 z y- \pi^2 z^2
\end{equation}
with  particular solution
\begin{equation}\label{eq11}
y_{p}(z)=i\pi z. \\
\end{equation}
Since $v(z)=c_1 e^{\int y_pdz}$, then a particular solution for (\ref{eq2}) is
\begin{equation}\label{eq13}
\begin{array}{l}
v_{p}(z)=c_1e^{i\frac {\pi}{ 2} z^2}~,
\end{array}
\end{equation}
which can be also obtained from (\ref{eq5}) by setting  $c_2=i c_1$.
Thus,   the  particular solution of  (\ref{eq1}) for  ${\rm w}(0)=0$ is
\begin{equation}\label{eq14}
\begin{array}{l}
{\rm w}_{p}(z)=c_1 \int_0^z e^{i\frac {\pi}{ 2} s^2}ds~.\\
\end{array}
\end{equation}

To construct the  general solution of Riccati equation (\ref{eq10}) using any particular solution  $y_p$  we let
 \begin{equation}\label{mhG}
y(z)=y_p(z)+\frac{1}{u(z)}~,
\end{equation}
where $u$ satisfies  the linear equation
\begin{equation}\label{mh5}
u'+\left(\frac 1 z-2 y_p\right)u=1~.
\end{equation}
The solution of (\ref{mh5}) is
\begin{equation}\label{mh6}
u(z)=\frac{\gamma+\int_0^z\mu(s)  ds}{\mu(z)}~,
\end{equation}
where $\mu(z)$ is the integrating  factor
\begin{equation}\label{mh7}
\mu(z)=ze^{-2\int y_p(z)dz}~,
\end{equation}
which gives the general solution
\begin{equation}\label{mh8}
y_g(z)=y_p(z)+\frac{\mu(z)}{\gamma+\int_0^z\mu(s) ds}~,
\end{equation}
and $\gamma$ arbitrary.

We now use the particular solution given by (\ref{eq11}) to construct the linear equation in $u$  which becomes
\begin{equation}\label{es1}
u\rq{}+\left(\frac 1 z -2 i \pi z\right)u=1.
\end{equation}
By using the integrating factor
\begin{equation}\label{es2}
\mu(z)=z e^{-i \pi z^2}~,
\end{equation}
the general solution of Riccati equation (\ref{eq10}) is
\begin{equation}\label{es3}
y_g(z)=\pi z \left(i+\frac{2}{i +(2 \pi \gamma - i)e^{i \pi z^2}}\right)~.
\end{equation}
By redefining the constant $\gamma=\frac{i(\theta+1)}{2\pi}$, 
(\ref{es3}) takes the simpler  form
\begin{equation}\label{es4}
y_g(z)=i\pi z\left(\frac{\theta e^{i \pi z^2}-1}{\theta e^{i \pi z^2}+1}\right)~.
\end{equation}
Notice that for the limiting cases of $\theta \rightarrow 0$, and $\theta \rightarrow \infty $ then $y_g(z) \rightarrow -i \pi z$, and $y_g(z) \rightarrow i \pi z$ respectively.
When $\theta \rightarrow 1$,  and  $\theta \rightarrow -1$,  then $y_g(z) \rightarrow - \pi z \tan \left( \frac{\pi z^2}{2}\right)$, and  $y_g(z) \rightarrow  \pi z \cot \left( \frac{\pi z^2}{2}\right)$  respectively.

\begin{center}
\begin{figure}[h!] 
\includegraphics[width=1.05\textwidth]{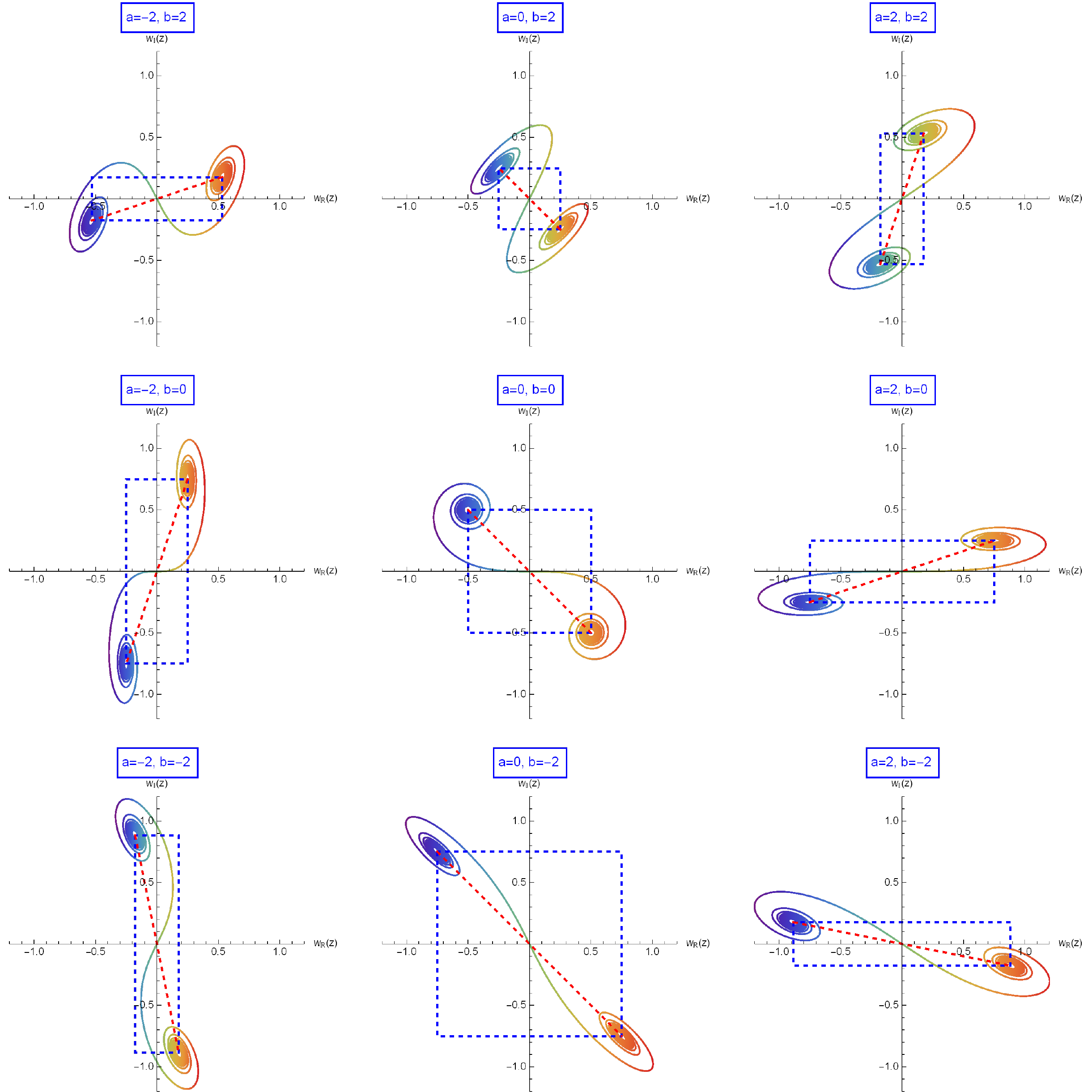} 
\caption{Argand plots for the Riccati deformed Cornu spirals for different values of $a$ and $b$. The $a=0, b=0$ case at the centre corresponds
to the supersymmetric partner equation (\ref{feq3}), while all the other cases correspond to equation (\ref{feq3bb}).}
\label{fig1}
\end{figure}
\end{center}

To find the general solution for ~(\ref{eq2}), we use $v_g(z)=Re^{\int y_g(z)dz}$ to obtain
\begin{equation}\label{es5}
v_g(z)=R\left( e^{-i\frac {\pi}{ 2} z^2}+\theta  e^{i\frac {\pi}{ 2} z^2}\right).
\end{equation}
By one integration, assuming ${\rm w}(0)=0$ and using Euler's formula, the deformed solution of (\ref{eq1}) is given by
\begin{equation}\label{es6}
{\rm w}_g(z)=R[(1+\theta) C(z)+i(-1+\theta)S(z)]~.
\end{equation}
By writing the solution as
\begin{equation}\label{es7}
{\rm w}_g(z)={\rm w}_{\mathcal R}(z)+i {\rm w}_{\mathcal I}(z)
\end{equation}
and letting  $\theta=a+i b$, we obtain
\begin{equation}\label{es8}
\begin{array}{l}
{\rm w}_{\mathcal R}(z)=R[(a+1)C(z)-bS(z)]~, \\
{\rm w}_{\mathcal I}(z)=R[bC(z)+(a-1)S(z)].
\end{array}
\end{equation}
Comparing (\ref{eq7}) with (\ref{es7}) and (\ref{es8}), one can see that we managed to replace the superposition constants $c_1$ and $c_2$ by the real and imaginary components of the parameter entering the general Riccati solution. This is not a trivial replacement because as we will see next one can disentangle an underlying supersymmetric structure of the solution expressed in this way.
We present the Argand plots $Y={\rm w}_{\mathcal I}(z)$, $X={\rm w}_{\mathcal R}(z)$ in  Fig.~\ref{fig1} for various parameters $a$ and $b$. All figures except $a=0,~b=0$ are scaled by the factor $R=\frac{1}{\sqrt{a^2+b^2}}$.

\section{Factorization of equation (\ref{eq2}) and supersymmetric approach}
We will now demonstrate the supersymmetric features of the solution in which the complex Riccati parameter is used.

\begin{center}
\begin{figure}[h!] 
\includegraphics[width=1.05\textwidth]{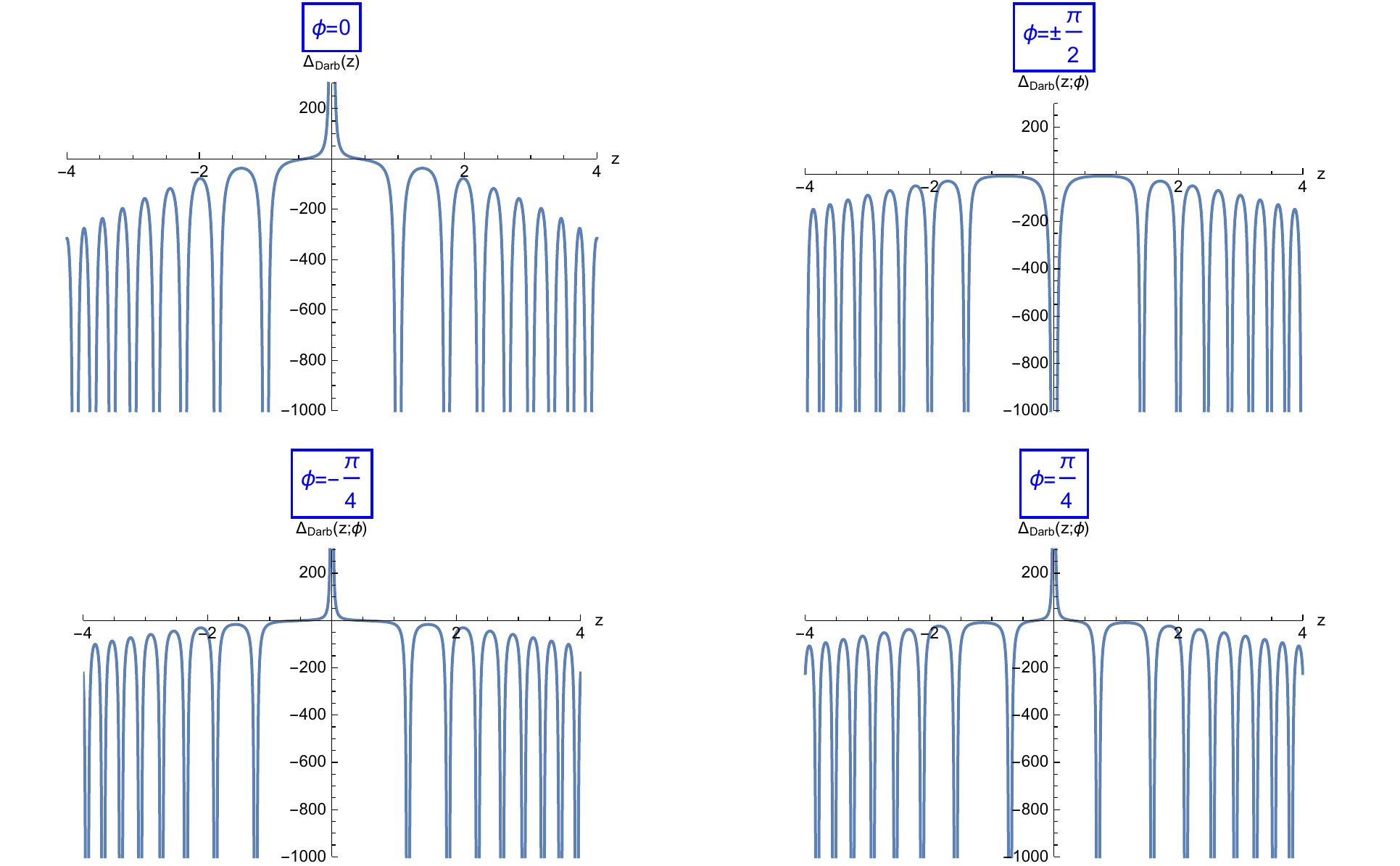}
\caption{Darboux distortion of (\ref{eq2}) for various phases $\phi$. The Darboux distortion of the supersymmetric partner of the Cornu spiral corresponds to $\phi=0$. The other cases correspond to members of the parametric deformed family of spirals having the same supersymmetric partner.}
\label{fig2}
\end{figure}
\end{center}

Equation~(\ref{eq2}) can be written in the factorized form $A^-A^+v=0$ using the differential operators given by
\begin{equation}
\begin{array}{l}
A^+=\frac{1}{\sqrt{z}}\frac{d}{dz}+\pi\sqrt{z}\tan \frac{\pi z^2}{2}\\
A^-=\frac{d}{dz}\frac{1}{\sqrt{z}}-\pi\sqrt{z}\tan \frac{\pi z^2}{2}~.
\end{array}
\end{equation}
Proceeding like in supersymmetric quantum mechanics, the supersymmetric partner equation of (\ref{eq2}) is
\begin{eqnarray}\label{feq3}
A^+A^-\Psi\equiv  
\Psi\rq{}\rq{}-\frac 1 z \Psi \rq{}+\left(\pi^2 z^2+\Delta_{{\rm Darb}}(z)\right) \Psi =0~.
\end{eqnarray}
The extra term in (\ref{feq3}) with respect to (\ref{eq2}) given by
\begin{equation}\label{feq3c}
\Delta_{{\rm Darb}}(z)=-2\pi^2z^2+\frac{3}{4z^2}-\pi \tan \frac{\pi z^2}{2}-2\pi^2 z^2\tan ^2\frac{\pi z^2}{2}
\end{equation}
is the Darboux distortion of (\ref{eq2}), and is presented in the first plot of Fig.~\ref{fig2}.

\begin{center}
\begin{figure}[h!] 
\includegraphics[width=0.55\textwidth]{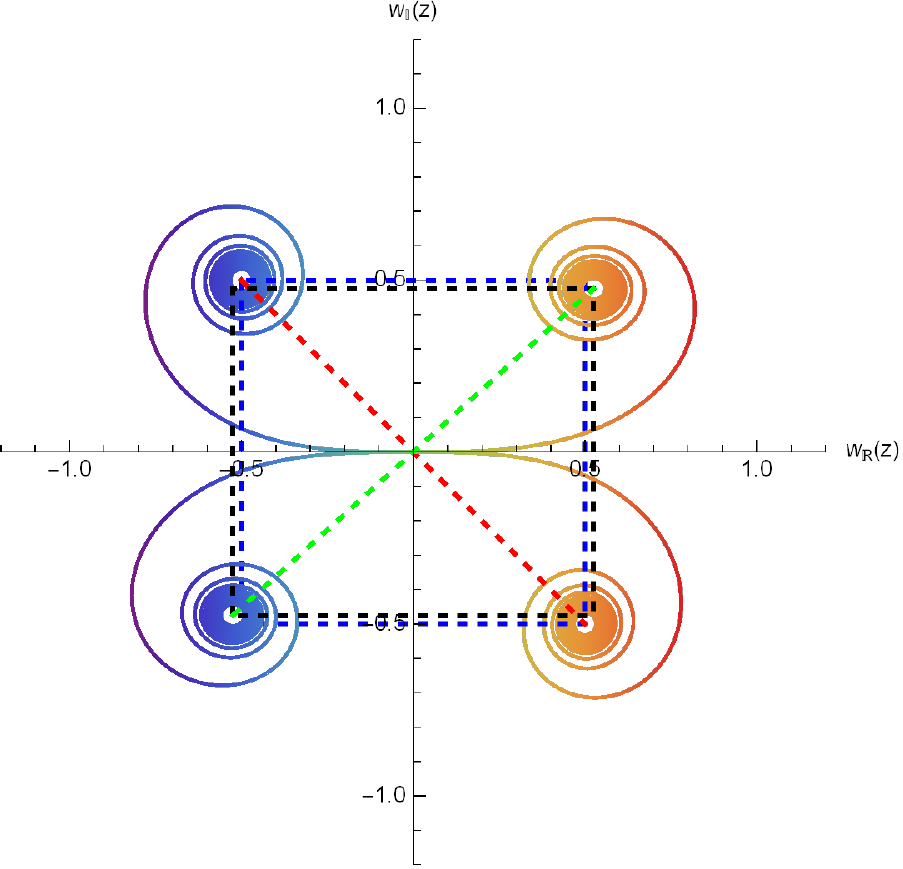}
\caption{The Cornu supersymmetric partner spiral, $a=0, b=0$, from the centre of Fig.~(\ref{fig1}) and the parametric Cornu spiral for $a=10, b=0$ which is already very close to the standard Cornu spiral. Notice also that in the limit $a\rightarrow \infty$ the supersymmetric partner spiral is the image of the standard spiral under real axis reflection.}
\label{fig3}
\end{figure}
\end{center}

To find out what second-order linear ODE corresponds  to the deformed Cornu spirals, we first write the general Riccati solution (\ref{es4}) in the trigonometric form
\begin{equation}\label{es5b}
y_g(z)=-\pi z\tan \left (\frac{\pi z^2}{2}+\phi \right)~, \qquad \theta = \frac 1 R e^{i\phi}~,
\end{equation}
and simply substitute it instead of the particular Riccati solution in the factorization (\ref{feq3})
\begin{equation}\label{factgen}
\left[\frac{1}{\sqrt{z}}\frac{d}{dz}-\frac{y_g(z)}{\sqrt{z}}
\right]
\left[\frac{d}{dz}\frac{1}{\sqrt{z}}+\frac{y_g(z)}{\sqrt{z}}
\right]
\widetilde{\Psi}=0~.
\end{equation}
One obtains the equation
\begin{equation}\label{feq3bb}
\widetilde{\Psi}\rq{}\rq{}-\frac 1 z \widetilde{\Psi} \rq{}+\left[\pi^2 z^2+\Delta_{{\rm Darb}}(z;\phi)\right]\widetilde{\Psi} =0~,
\end{equation}
with the Darboux distortion depending parametrically on the phase shift $\phi$
\begin{eqnarray}\label{Ddp}
&\Delta_{{\rm Darb}}(z;\phi)=-2\pi^2z^2+\frac{3}{4z^2}-\pi \tan  \left (\frac{\pi z^2}{2}+\phi \right)\nonumber\\
& -2\pi^2 z^2\tan ^2 \left (\frac{\pi z^2}{2}+\phi \right)~.\,
\end{eqnarray}
To find the general solution to (\ref{feq3bb}), we let $A^-\widetilde \Psi=\widetilde  \Phi$, thus the  homogenous equation
$A^+\widetilde \Phi=0$ has solution
\begin{equation}
\widetilde \Phi(z; \phi)=b_1 \cos  \left (\frac{\pi z^2}{2}+\phi \right).
\end{equation}
By solving the  nonhomogeneous equation $A^-\widetilde \Psi=\widetilde \Phi$ we obtain the general solution of (\ref{feq3bb})
\begin{align}\label{f10}
&\widetilde \Psi(z; \phi)=\nonumber \\
&\frac{\sqrt z}{4}\frac{b_2+2 b_1 z+\sqrt 2 b_1\left[\cos 2 \phi~
C(\sqrt 2 z)-\sin 2 \phi ~ S(\sqrt 2 z) \right]}{\cos \left (\frac{\pi z^2}{2}+\phi \right)}.
\end{align}
Denoting
\begin{equation}
\widetilde C(z; \phi) =\int_{0}^{z} \cos ( \pi s^2+2 \phi) ds
\end{equation}
and choosing the arbitrary constants to be $b_1=2,b_2=0$,  (\ref{f10}) takes the compact form
\begin{equation}\label{f12}
\widetilde \Psi(z; \phi)=\sqrt z\frac{z+\widetilde C(z;\phi) }{\cos \left (\frac{\pi z^2}{2}+\phi \right)}.
\end{equation}
In Fig.~(\ref{fig2}), we display various cases of the parametric Darboux distortions $\Delta_{{\rm Darb}}(z;\phi)$ of the deformed Cornu spirals
presented in Fig.~(\ref{fig1}). We notice the negative parabolic envelope as given by the first term in (\ref{Ddp}) together with the singularities due to the terms containing the tangents for nonzero $z$. The singularities at the origin are due to the $1/z^2$ term except for the cases $\phi=\pm \pi/2$ when the dominant contribution comes from the cotangent terms. For these values of phase the Darboux distortion simplifies to
\begin{equation}
\begin{array}{ll}
&\Delta_{{\rm Darb}}(z;0)=\frac{3}{4z^2}-\frac{\pi\left(\sin \pi z^2+4 \pi z^2\right)}{\cos \pi z^2+1},\nonumber \\
&\Delta_{{\rm Darb}}(z;\pm\pi/4)=\frac{3}{4z^2}+\frac{\pi\left(\cos \pi z^2\pm 4 \pi z^2\right)}{\sin \pi z^2\mp 1},\nonumber \\
&\Delta_{{\rm Darb}}(z;\pm\pi/2)=\frac{3}{4z^2}-\frac {\pi \left(\sin \pi z^2-4 \pi z^2\right)}{\cos \pi z^2-1}.\nonumber
\end{array}
\end{equation}
The factorization patterns discussed here unravel the Darboux origin of this deformation, which is a counterpart of the same construction in supersymmetric quantum mechanics, where parametric families of supersymmetric isospectral potentials have been obtained with the property that all the members of those families have the same supersymmetric partner potential \cite{M84,RMC1,R98}. The initial potential and its supersymmetric partner are reproduced for extremal values of the parameters. In the case of the Cornu spiral, our parametrization has been chosen such that when $a$ and $b$ are nought the supersymmetric partner spiral is obtained, whereas the standard Cornu spiral is obtained when $a\rightarrow \infty$ and $b=0$. This is graphically demonstrated in Fig.~(\ref{fig3}) where even for the rather small value of $a=10$ and $b=0$, the spiral is very close to the standard one as known from textbooks \cite{Fowles}.

\section{Conclusion}

A parametric deformation of the Cornu spiral has been introduced based on the usage of the corresponding general Riccati solution instead of the particular solution. Geometrically, the origin of this kind of deformation lies in the two independent shifts, $a$ and $b$, along the two orthogonal axes of the plane in which the spiral is plotted. These shifts can generate not only the deformation of the rolls of the spiral but also its global rotation as seen in the plots.
Foreseen applications are in the same range as those of the standard Cornu spiral.

\bigskip



\end{document}